\documentclass[journal,10pt,a4paper,twocolumn,twoside]{IEEEtran}

\IEEEoverridecommandlockouts
\usepackage{cite}
\usepackage{amsmath,amssymb,amsfonts}
\usepackage{algorithm}
\usepackage{algorithmicx}
\usepackage{algpseudocode}
\usepackage{setspace}
\usepackage{bm}
\usepackage{graphicx}
\usepackage{textcomp}
\usepackage{xcolor}
\usepackage{tikz}
\usetikzlibrary{quantikz2}
\usepackage{float}
\usepackage{subcaption}
\usepackage[T1]{fontenc}
\usepackage{verbatim}
\usepackage{hyperref}
\usepackage{makecell}
\usepackage{booktabs}
\usepackage{array} 
\usepackage{amsthm}
\usepackage{graphbox} 
\usepackage{adjustbox}
\usepackage{balance}
\usepackage{multirow}
\usepackage[capitalize]{cleveref}
\usepackage{footnote}
\usepackage{bm}
\usepackage{threeparttable}
\usepackage{caption}
\usepackage{lipsum}
\usepackage{nicematrix}
\usepackage[usestackEOL]{stackengine}

\newtheorem{lem}{Lemma}

\theoremstyle{remark}

\theoremstyle{definition}
\newtheorem{defin}{Definition}

\newtheorem*{remark}{Remark}

\newtheorem*{main}{Main idea}
\newtheorem*{example}{Example}

\newcommand{\eqdef}{\stackrel{\triangle}{=}}

\begin{document}

\title{Beyond Traditional Quantum Routing
}

\author{Si-Yi Chen, Angela~Sara~Cacciapuoti,~\IEEEmembership{Senior~Member,~IEEE}, and Marcello~Caleffi,~\IEEEmembership{Senior~Member,~IEEE}
    \thanks{This work has been accepted in the Proceedings of IEEE QCE’25.}
    \thanks{The authors are with the Quantum Internet Research Group \href{www.quantuminternet.it}{(www.QuantumInternet.it)}, University of Naples Federico II, Italy. }
    \thanks{Corresponding author: Angela Sara Cacciapuoti (e-mail: angelasara.cacciapuoti@unina.it).}
    \thanks{This work has been funded by the European Union under Horizon Europe ERC-CoG grant QNattyNet, n.101169850. Views and opinions expressed are however those of the author(s) only and do not necessarily reflect those of the European Union or the European Research Council Executive Agency. Neither the European Union nor the granting authority can be held responsible for them.}
    }
\maketitle

\begin{abstract}
Existing quantum routing implicitly mimics classical routing principles, with finding the ``best'' path (aka pathfinding), according to a selected routing metric, as a core mechanism for establishing end-to-end entanglement. However, optimal pathfinding is computationally intensive, particularly in complex topologies. In this paper, we propose a novel approach to quantum routing, which avoids the inherent overhead of conventional quantum pathfinding, by establishing directly entanglement between remote nodes. Our approach exploits graph complement strategies. It allows to improve the flexibility and efficiency of quantum networks, by paving the way for more practical quantum communication infrastructures.
\end{abstract}


\section{Introduction}
\label{sec:1}
Enabling end-to-end entanglement between remote network nodes is often considered as a primary task in quantum networks. Research in this field has extensively drawn inspiration from traditional Internet routing principles, such as utilizing quantum repeaters to extend end-to-end entanglement\cite{Cal-17,ShoQia-20,AliChe-22,LiuLiCai-24,LiuLiWang-25} along selected paths. In this context, finding the “best” path (aka pathfinding), according to a selected routing metric, is considered as the core
mechanism for establishing end-to-end entanglement -- also referred in the following as virtual link. 

However, and as recently shown in \cite{CacIllCal-23}, quantum routing is much more than merely discovering paths toward destination nodes. It should be able to track entanglement resources within the network \cite{CacIllCal-23}. To this aim, multipartite entanglement significantly enhances the connectivity features of entanglement-based networks~\cite{CheIllCac-24-QCE,PirDur-19,LiXueLi-23, CheIllCac-25, MazCalCac-24,MorDur-24,WenJosSte-18}. And these features are, in turn, strongly influenced by the specific class of selected multipartite entanglement state. Research has shown that graph states constitute an interesting class of multipartite entanglement for quantum network applications \cite{HeiDurEis-06}.  

In this paper, we exploit graph states for proposing a radically new approach to quantum routing, by leveraging the entanglement-enabled connectivity\cite{CacIllCal-23}.

\begin{table}[!t]
\caption{Traditional Quantum Routing (TQR) VS Proposed Approach}
\label{tab:01}
\centering
\footnotesize
\setlength{\tabcolsep}{4pt}
\renewcommand{\arraystretch}{2}
\begin{threeparttable}[t]
\begin{tabular}{c|c|c}
    \toprule
    \toprule
    & \textbf{TQR} & \textbf{Proposed Approach} \\
    \midrule
    \textbf{Key Operation} & Path Selection & Graph Manipulation \\
    \midrule
    \textbf{Entanglement Resource} & EPR Pairs & Graph States \\
    \midrule
    \textbf{Entanglement Distribution} & Reactive & Proactive \\
    \midrule
    \textbf{Key Tool} & Entanglement Swapping & Pauli-Measurement \\
    \bottomrule
\end{tabular}
\end{threeparttable}

\end{table}

Specifically, we propose a graph complement strategy for addressing the quantum routing problem, to directly and simultaneously neighbor all remote source-destination pairs within the artificial topology, activated by the multipartite entanglement. Then we exploit the proposed strategy for the communications among nodes belonging to different Quantum Local Area Networks (QLANs), namely for Inter-QLAN communications.

Our findings, although preliminary, pave the way for establishing end-to-end entanglement between pairs of nodes by saving in delay and overhead. In fact, the proposed approach avoids the heavy coordination and signaling traffic loads required by traditional quantum routing approaches.

It should be emphasized that the differences between traditional quantum routing and the proposed graph complement strategy span multiple aspects, as summarized in Table~\ref{tab:01} and elaborated in Section~\ref{sec:2}.


\section{Graph Complement VS\\ Traditional Quantum Routing}
\label{sec:2}

\begin{figure*}[t]
    \centering    
    \includegraphics[width=\textwidth]{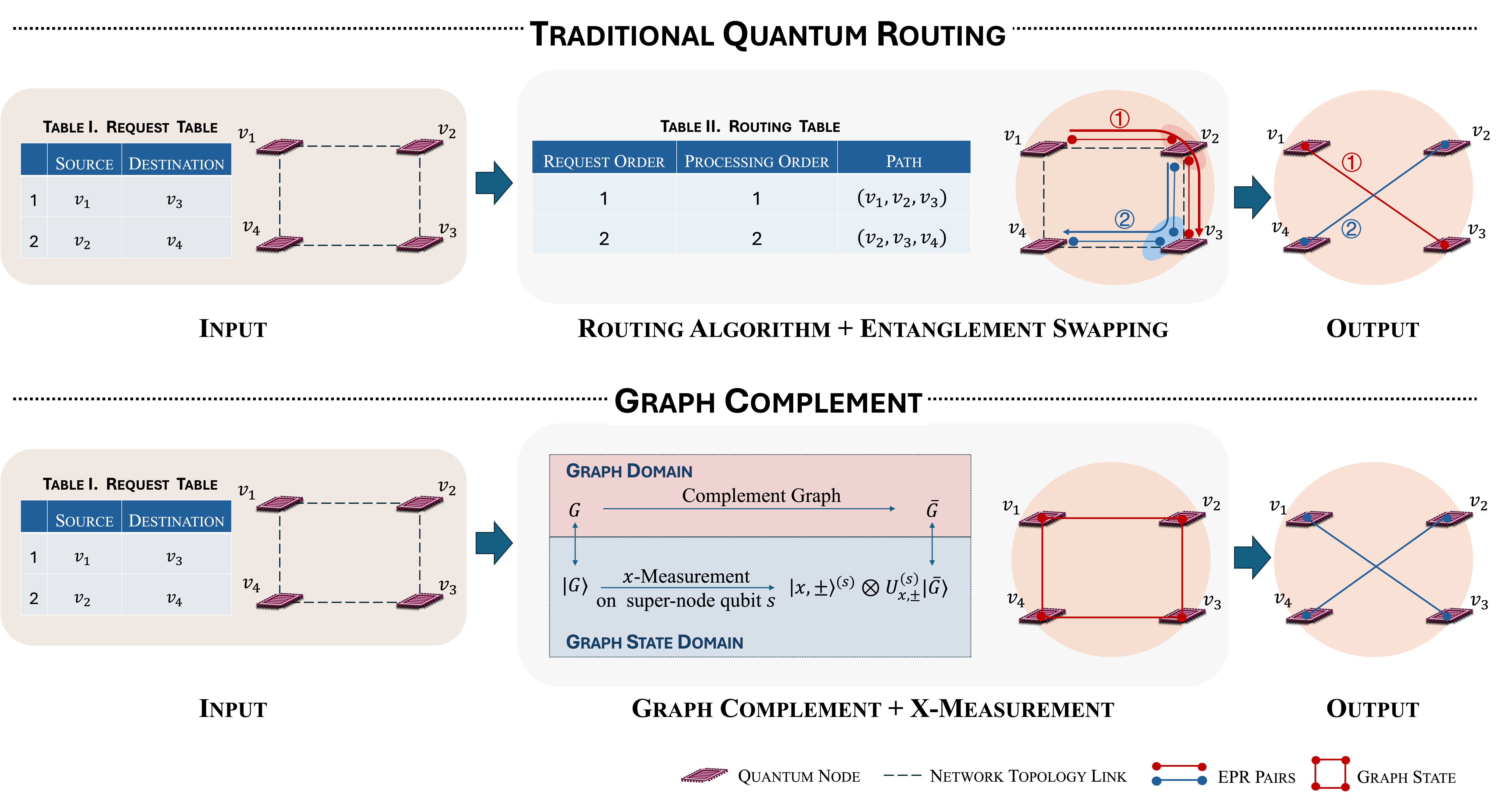}
    \caption{Traditional Quantum Routing VS Graph Complement.}
    \label{fig:01}
    \hrulefill
\end{figure*}

\subsection{Traditional Quantum Routing}
\label{sec:2.1}

As aforementioned, in traditional quantum routing for entangling source with the destination, the very same problem underlying classical routing, i.e., discovering a path toward the destination, must have been solved beforehand \cite{CacIllCal-23}. This implies that entanglement generation and distribution process is performed \textit{reactively}, meaning that it is triggered only after the path has been discovered. Consequently, it depends on the specificity of the path discovery functionality.
As illustrated in Fig.~\ref{fig:01}, traditional quantum routing takes a routing request as input, which consists of a series of source-destination pairs. The primary task is to identify a path from source to destination within the network topology, scheduling each repeater along the path to generate entanglement with its neighboring nodes. For example, to fulfill the first request depicted in Fig.~\ref{fig:01}, the path $(v_1, v_2, v_3)$ is discovered and selected to establish a virtual link between the two endpoints, where $v_2$ functions as a quantum repeater. Subsequently, entanglement swapping operations (denoted with light red background in $v_2$, and light blue background in $v_3$) are performed to establish an end-to-end virtual links between sources and destinations. 

It is worthwhile to mention that \textit{communication qubits}\cite{CalAmoFer-24,rfc9340}, i.e., qubits reserved for the generation and distribution of the entangled states, play a crucial role in quantum routing. Indeed, each node must have at least one communication qubit, and a larger number of communication qubits can significantly improve the end-to-end entanglement rate, with an obvious positive effect on the number of requests that can be accommodated in parallel. Conversely, if the number of communication qubits is limited to one, establishing simultaneously multiple end-to-end entanglement across multiple links becomes unfeasible. For example and as shown in Fig.~\ref{fig:01}, when repeaters $v_2$, $v_3$ are limited to have only two communication qubits each, different input requests can only be processed in a single-threaded manner, being the parallel execution of tasks forbidden.

\subsection{Graph Complement Strategy}
\label{sec:2.2}

An alternative approach to quantum routing consists in exploiting the entanglement-enabled connectivity, activated by multipartite states, to establish end-to-end entanglement between remote nodes. By utilizing graph states as communication resources, this approach simplifies to properly manipulate the associate graph\cite{CheIllCac-25}.

Specifically, graph complement on the graph associated to the original graph state creates edges between all previously non-adjacent nodes, by transforming them from remote to adjacent nodes. We envision to built the routing strategy on this manipulation, since it aligns closely with the requirements of establishing end-to-end entanglement between remote nodes. Remarkably, in the graph state domain, the graph complement has a corresponding measure operation, implemented through $X$-measurement, as depicted in Fig.~\ref{fig:01}. 

To further intuitively grasp the differences between traditional quantum routing and graph complement strategy introduced so far, we can emphasize that the graph complement strategy is able (differently from the traditional quantum routing approach) to accommodate multiple requests in parallel, even if each network node is constrained to have only one communication qubit. For instance, as shown in Fig. ~\ref{fig:01}, although each node has only one communication qubit, the proposed strategy is able to serve both the requests simultaneously. This advantage becomes particularly significant in complex inter-domain networks.

To the best of our knowledge, this is the first study to tackle the challenge of establishing remote end-to-end entanglement without relying on conventional pathfinding mechanisms.

\section{Problem Statement}
\label{sec:3}

In this section, we first introduce some preliminaries used in the remaining part of the paper, and then we provide a formal definition of the research problem.

\begin{figure*}[t]
    \centering    \includegraphics[width=\textwidth]{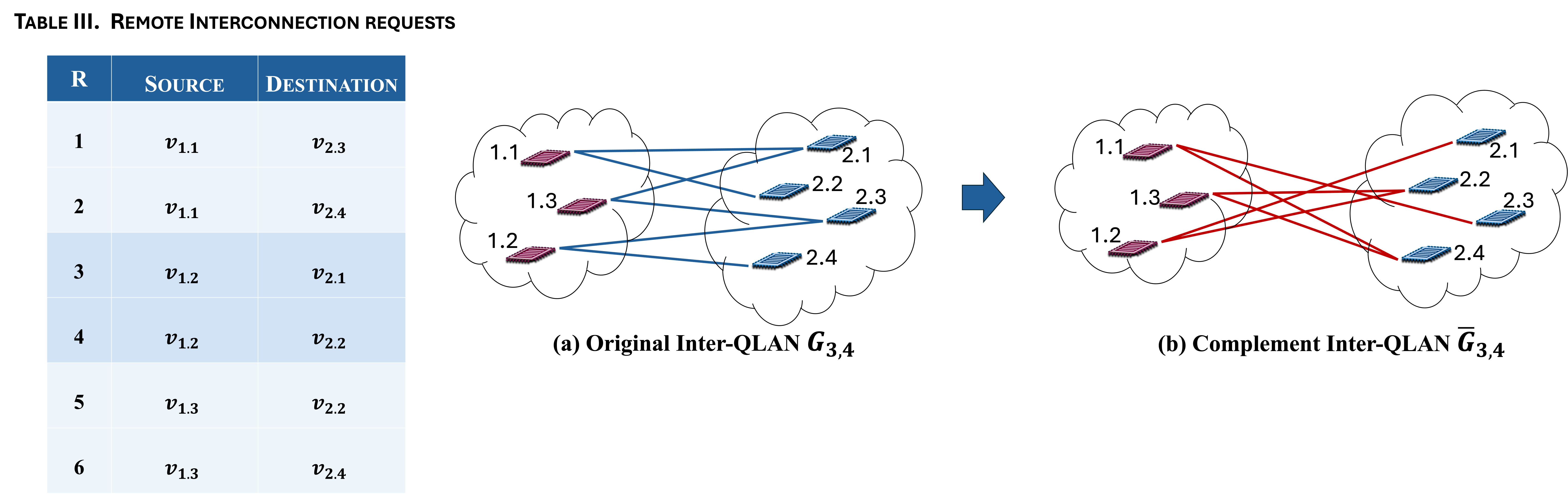}
    \caption{Pictorial illustration of research problem and main idea. Consider an Inter-QLAN network $\ket{G_{3, 4}}$ with corresponding graph $G_{3,4}=(V_1, V_2, E)$ as shown in Fig.~\ref{fig:02}(a). Each client node $v_i \in V_1$ denote with red chip, while $v_j \in V_2$ denoted with blue chip. A set of remote interconnection requests $R$ is listed on the left Tab.~III. Instead of searching for paths to satisfy each request, we transform the original Inter-QLAN network into its complement one, where all the requested node pairs in $R$ become directly connected.
    }
    \label{fig:02}
    \hrulefill
\end{figure*}

\subsection{Preliminaries}
\label{sec:3.1}

We assume a graph state $\ket{G}$, distributed between two QLANs. This multipartite entanglement state enables an artificial topology built upon the physical one \cite{IllCalMan-22,CheCacCal-25}. In the following, we provide some definition with reference to this artificial topology.

\begin{defin}[\textbf{Inter-Links}]
    \label{def:x01}
    Given two QLANs $V_1, V_2$, the Inter-Links is the edge $E_{i,j}$ connected node $v_i \in V_1$ in one QLAN with another node $v_j \in V_2$ in the other QLAN. Formally:
    \begin{equation}
        \label{eq:def:x01.1}
        E_{i,j}  \eqdef (v_i \in V_1 , v_j \in V_2). 
    \end{equation}
\end{defin}

Notably, $v_i, v_j$\footnote{\label{ft:01}For the sake of clarity, we also use $v_{1.i},v_{2.j}$ to clarify the node $v_i \in V_1, v_j \in V_2$. Clearly, a inter-link $E_{i,j}$ and a source-destination pairs $(v_{1.i}, v_{2.j})$ can be described equivalently.} in $E_{i,j}$ are strict in different QLANs, i.e., the order of the subscripts of $E_{i,j}$ implies the QLAN to which $v_i, v_j$ respectively belongs. Hence in this paper, $E_{i,j} \neq E_{j,i}$.  
Furthermore, we denote with $ N(v_i) \subseteq V_2$ the set of nodes in QLAN $V_2$ connected to $v_i \in V_1$.

Further, in the following, we use notation $E$ to denote the set of inter-links among two QLANs $V_1, V_2$. Formally:
\begin{equation}
        \label{eq:def:x01.2}
        E  \eqdef \bigcup_{ v_i \in V_1 , v_j \in V_2 }  E_{ij}, 
\end{equation}
with $E_{i,j}$ defined in \eqref{eq:def:x01.1}.

\begin{defin}[\textbf{Inter-QLAN}]
    \label{def:x02}
    Given two QLANs $V_1, V_2$, if there exists at least one inter-Link between them, we say that the two QLANs are \textit{interactive}, and the network constituted by all the nodes of QLANs with their inter-links is defined as Inter-QLAN. The inter-QLAN is described by the graph state $\ket{G_{n_1,n_2}}$, with associated graph $G_{n_1,n_2}$ given by:
    \begin{equation}
        \label{eq:def:x02}
        G_{n_1,n_2}  \eqdef (V_1, V_2, E), 
    \end{equation}
    with $|V_i|=n_i$ and $E$ defined in~\eqref{eq:def:x01.2}.
\end{defin}

Accordingly, the Inter-QLAN network concept emphasizes artificial links between QLANs. 

\begin{defin}[\textbf{Complement Inter-Links}]
    \label{def:x03}
    Given an Inter-QLAN $\ket{G_{n_1,n_2}}$ with corresponding graph $G_{n_1,n_2}=(V_1, V_2, E)$, the complement Inter-links of the arbitrary vertex $v_i \in V_i$ is the set $\bar{E}(v_i)$ of edges connecting $v_i$ with its remote nodes in a different QLAN $V_{\bar{i}}$. Formally:
    \begin{equation}        
        \label{eq:def:x03.1}
        \bar{E}(v_i)  \eqdef \bigcup_{v_j \in V_{\bar{i}}} (v_i \in V_i, v_j \in \overline{N}(v_i) ),    
    \end{equation} 
with $\overline{N}(v_i) \eqdef V_{\bar{i}} \setminus N(v_i)$.
\end{defin}
In the following, we use the symbol $\bar{E}$ to denote the set of complement Inter-links among the different QLANs $V_1, V_2$. Formally:
\begin{equation}
        \label{eq:def:x03.3}
         \bar{E}  \eqdef \bigcup_{ v_i \in V_1} \bar{E}(v_i).
\end{equation}

\begin{defin}[\textbf{Complement Inter-QLAN}]
    \label{def:x04}
    Given an Inter-QLAN $\ket{G_{n_1,n_2}}$ with corresponding graph $G_{n_1,n_2}=(V_1, V_2, E)$, its complement Inter-QLAN $\Bar{G}_{n_1,n_2}$ consists of the same vertices belonging to $G_{n_1,n_2}$, while the edge set is the set of complement Inter-links:
    \begin{equation}
        \label{eq:def:x04}
        \bar{G}_{n_1,n_2} \eqdef  \big ( V_1, V_2, \bar{E} \big),
    \end{equation}
    with $|V_i|=n_i$ and complement Inter-links set $\bar{E}$ defined in~\eqref{eq:def:x03.3}.
\end{defin}

\subsection{Problem statement}
\label{sec:3.2}

With above preliminaries, we define the routing problem for Inter-QLAN communications as follows.

\begin{main}
 Rather than relying on optimal pathfinding as in traditional quantum routing approaches, we directly switch the original Inter-QLAN network to its complement one, which simultaneously enables each pair of remote nodes in different QLAN to become neighborhoods. To realize such a complement Inter-QLAN, we propose two lemmas, i.e. Lem.~\ref{lem:x01} and~\ref{lem:x02} that formalize the conditions and mechanisms required for this transformation.
\end{main}

For the sake of clarity, we provide a pictorial illustration of the research problem in Fig.\ref{fig:02}. To explicitly demonstrate the remote interconnection requests, we list all possible remote requests in the accompanying table. In practical scenarios, the network may initiate a subset of these requests, corresponding to part of the inter-links illustrated in Fig.\ref{fig:02} (b).

\section{Beyond traditional quantum routing: \\ A Graph Complement strategy}
\label{sec:4}

\begin{figure*}[t]
    \centering
    \small 
    \begin{subfigure}[b]{\textwidth}
        \begin{subfigure}[t]{0.5\textwidth}
            \centering
            \includegraphics[width=0.75\textwidth]{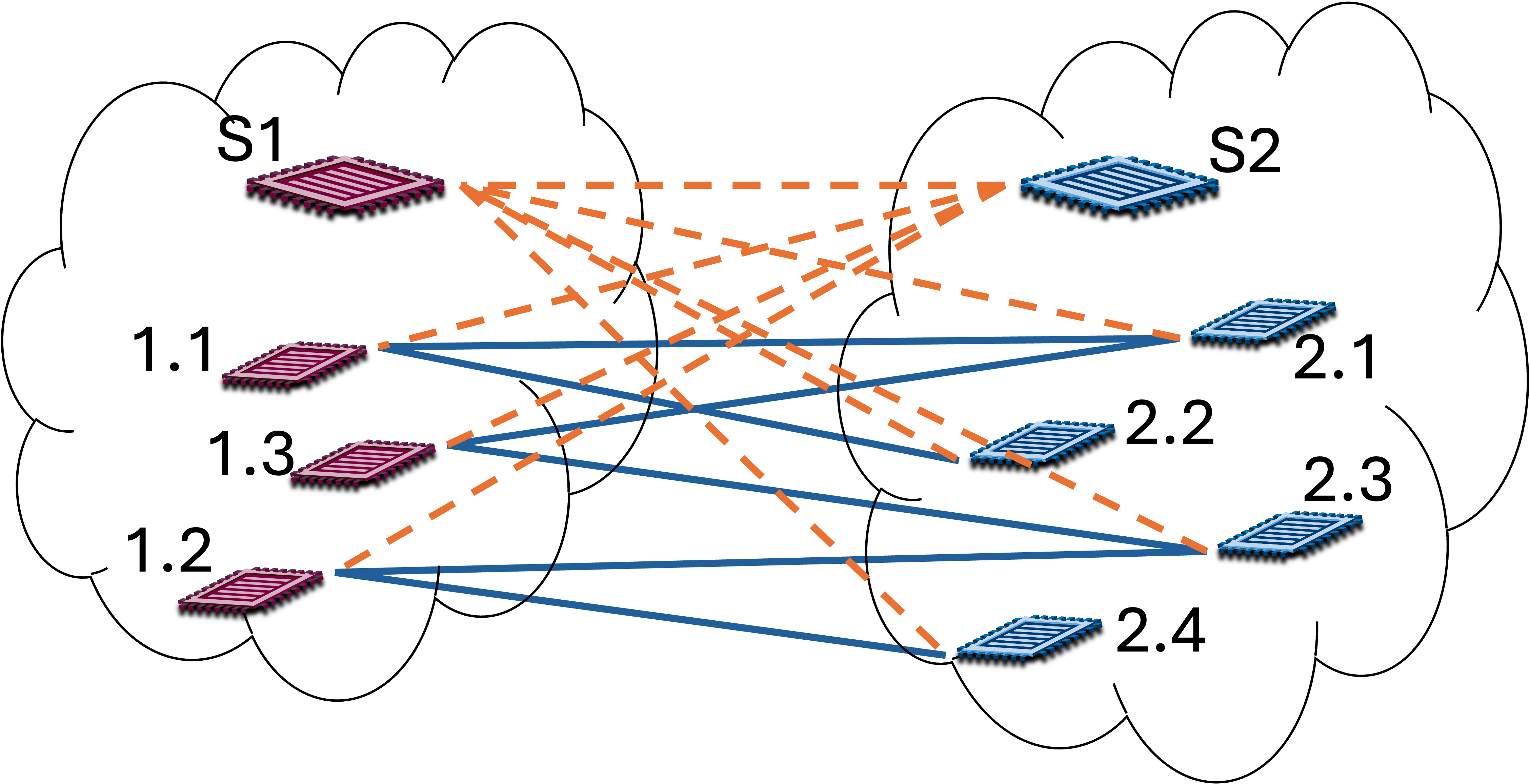}
            \subcaption{An Inter-QLAN $G_{4,5}$ satisfying Lem.~\ref{lem:x01}}
            \label{fig:03.a}
        \end{subfigure}
        \hspace{6pt}
        \begin{subfigure}[t]{0.5\textwidth}
            \centering
            \includegraphics[width=0.75\textwidth]{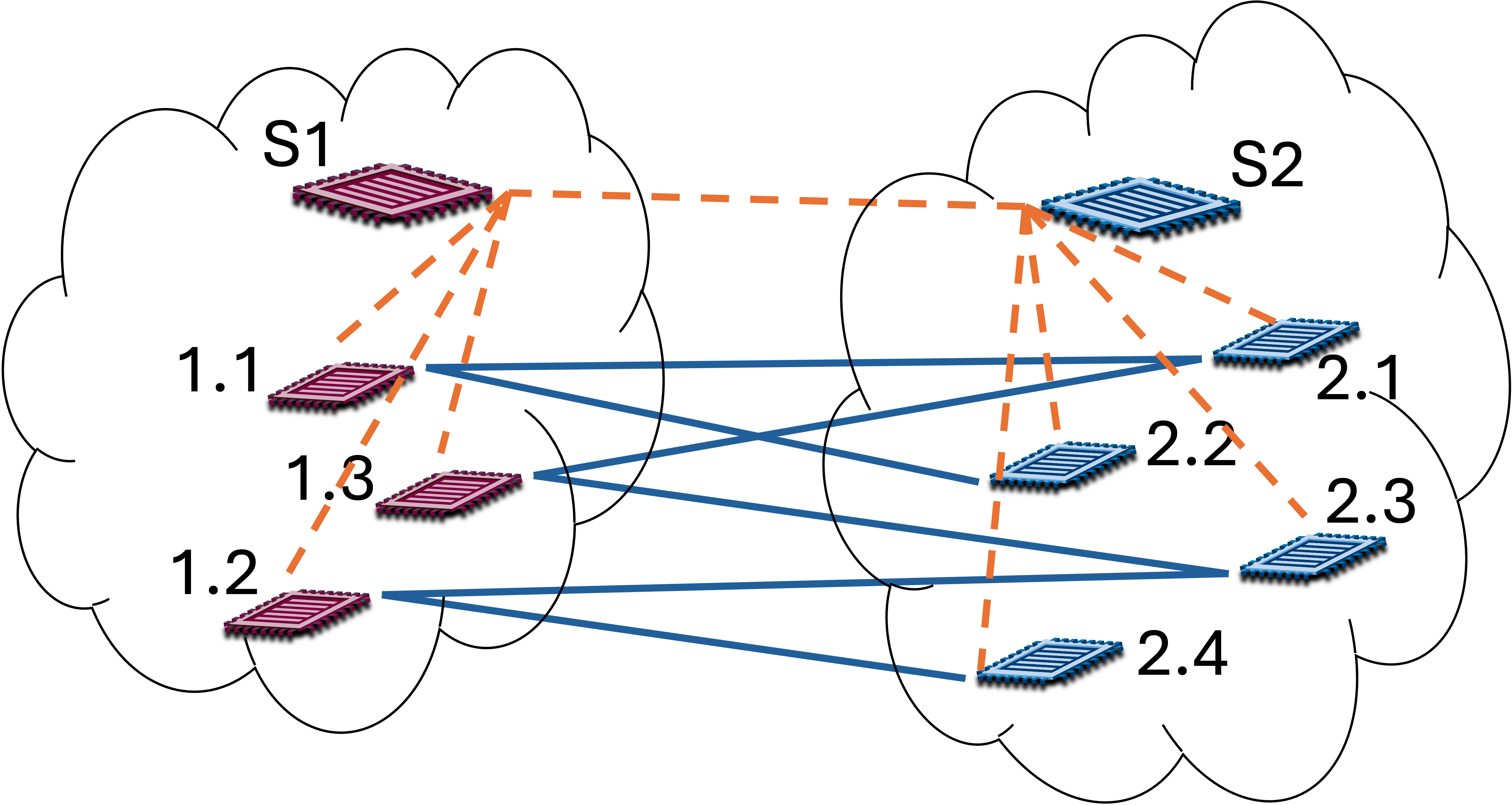}
            \subcaption{An Inter-QLAN-like $\tilde{G}_{4,5}$ satisfying Lem.~\ref{lem:x02}}
            \label{fig:03.b}
        \end{subfigure}
    \end{subfigure}
    \caption{Examples of Inter-QLAN and Inter-QLAN-like structures satisfying Lem.\ref{lem:x01} and Lem.\ref{lem:x02}. The Inter-QLAN $G_{4,5}$ in Fig.\ref{fig:03.a} and the Inter-QLAN-like structure $\tilde{G}_{4,5}$ in Fig.~\ref{fig:03.b} are both derived from the original Inter-QLAN $G_{3,4}$ in Fig.\ref{fig:02}(a) by introducing two connected super-nodes, $s_1$ and $s_2$ in different QLANs. In $G_{4,5}$, the super-nodes are connected to all client nodes in the opposite QLAN, whereas in $\tilde{G}_{4,5}$, the super-nodes are connected to all client nodes in the local QLAN. }
    \label{fig:03}
    \hrulefill
\end{figure*}

\subsection{Complement Inter-QLAN}
\label{sec:4.1}
To realize the graph complement, we derive two lemmas that enable its systematic construction. To this aim, we adopt as original multipartite entanglement resource, shared between the two QLANs, a bipartite graph state \cite{CheIllCac-25}. The structure of a bipartite graph state is such that there exists one node in each QLAN fully connected with the nodes of the other QLAN. We refer to this type of node as super-node, to distinguish it from plain nodes (aka not fully connected) referred to as clients. 

\begin{lem}
\label{lem:x01}
    Given a bipartite graph $\ket{G_{n_1+1,n_2+1}}$ shared in an Inter-QLAN network $G_{n_1+1,n_2+1}$ such that: 
    \begin{itemize}
    \item [(1)] Each client-node $v_i \in V_i$ in one QLAN is artificial connected
        \begin{itemize}
            \item [$\cdot$] with the client-nodes $N(v_i) \subseteq V_{\bar{i}}$ in the other QLAN;
            \item [$\cdot$] and with the super-node $s_{\bar{i}}$ in the other QLAN;
        \end{itemize}
    \item [(2)] Super-nodes ${s_1,s_2}$ in different QLANs are connected by one inter-link $(s_1, s_2)$.
    \end{itemize}
    If the bipartite graph satisfies the above conditions, it allows to obtain a $(n_1+n_2)$-qubit bipartite graph state $\ket{\bar{G}_{n_1,n_2}}$, distributed among a complement Inter-QLAN $\bar{G}_{n_1,n_2}$, connecting each pair of complement inter-links among client-nodes.
    Formally:
    \begin{align}
        \label{eq:lem:x01}
        &\ket{G_{n_1+1,n_2+1}} \xrightarrow{\text{LOCC}} \ket{\bar{G}_{n_1,n_2}},  \\ \nonumber
        \text{with} \;&G_{n_1+1,n_2+1} = ( V_1 \cup \{s_1\}, V_2 \cup \{s_2\},  \\ \nonumber
        & E \cup (s_1,s_2) \cup \big( \bigcup_{ \substack{ v_i \in V_1} } (v_i, s_2) \big) \cup
        \big( \bigcup_{v_i \in V_2} (v_i, s_1) \big)
    \end{align}
    with $V_1, V_2, E$ defined in $G_{n_1,n_2}$ in~\eqref{eq:def:x02}, and $\bar{G}_{n_1,n_2}$ given in~\eqref{eq:def:x04}, respectively.   
    \begin{IEEEproof}
        Please refer to App.~\ref{app:lem:x01}. 
    \end{IEEEproof}
\end{lem}

\begin{lem}
\label{lem:x02}
    Given a bipartite graph $\ket{\tilde{G}_{n_1+1,n_2+1}}$ shared between $(n_1+n_2)$ client-nodes belonging to an Inter-QLAN network $G_{n_1,n_2}$ and two super-nodes $s_1,s_2$ in two different QLANs such that: 
    \begin{itemize}
    \item [(1)] Each client-node $v_i \in V_i$ in one QLAN is connected
        \begin{itemize}
            \item [$\cdot$] with client-nodes $N(v_i) \subseteq V_{\bar{i}}$ in the other QLAN;
            \item [$\cdot$] and with the super-node $s_{i}$ in the local QLAN;
        \end{itemize}
    \item [(2)] Super-nodes ${s_1,s_2}$ in different QLANs are connected by one inter-link $(s_1, s_2)$.
    \end{itemize}
    If the bipartite graph satisfies the above conditions, it allows to obtain a $(n_1+n_2)$-qubit bipartite graph state $\ket{\bar{G}_{n_1,n_2}}$, distributed among a complement Inter-QLAN $\bar{G}_{n_1,n_2}$, connecting each pair of complement inter-links among client-nodes.
    Formally:
    \begin{align}
        \label{eq:lem:x02}
        &\ket{\tilde{G}_{n_1+1,n_2+1}} \xrightarrow{\text{LOCC}} \ket{\bar{G}_{n_1,n_2}},  \\ \nonumber
        \text{with} \;&\tilde{G}_{n_1+1,n_2+1} = ( V_1 \cup \{s_1\}, V_2 \cup \{s_2\},  \\ \nonumber
        & E \cup (s_1,s_2) \cup \big( \bigcup_{ \substack{ v_i \in V_1} } (v_i, s_1) \big) \cup
        \big( \bigcup_{v_i \in V_2} (v_i, s_2) \big)
    \end{align}
    with $V_1, V_2, E$ defined in $G_{n_1,n_2}$ given in~\eqref{eq:def:x02}, and $\bar{G}_{n_1,n_2}$ given in~\eqref{eq:def:x04}, respectively.   
    \begin{IEEEproof}
        Please refer to App.~\ref{app:lem:x01}. 
    \end{IEEEproof}
\end{lem}

\begin{remark}
Notably, Lem.~\ref{lem:x01} and Lem.~\ref{lem:x02} introduces two cases of super-nodes to enable inter-QLAN complement conversion:\\
\textbf{Case I}: If in each QLAN there exists at least one node satisfying Lem.~\ref{lem:x01}, which is connected to all nodes in the other QLAN, such particular node will act as super-node in QLAN. \\
\textbf{Case II}: If not, we can add one local super-node satisfying Lem.~\ref{lem:x02}, which is connected to every client-node in local QLAN and connected with the super-node in the other QLAN. Since the Inter-QLAN network strictly allows connections only between nodes in different QLANs, while the super-node in Case II exhibits a mild connection abuse, we prefer the network in Case II as an \textit{Inter-QLAN-like} network.
\end{remark}

\begin{remark}
    Lemma~\ref{lem:x01} and Lemma~\ref{lem:x02} also supports partial complement Inter-QLANs network switch. If one or several nodes wish to retain the original inter-link rather than switching to its complement link, it is sufficient that the super-node of that QLAN does not connect to such nodes without affecting the conversion of the remaining portion of the Inter-QLANs network to its complement one.
\end{remark}

\section{Conclusion}
\label{sec:5}

In this work, we proposed a graph complement strategy as a novel alternative to traditional pathfinding-based quantum routing. By avoiding the need for computationally intensive path selection, our approach enables direct entanglement between remote nodes in a more efficient manner. This significantly reduces routing overhead and enhances the overall flexibility of quantum networks. Our findings, although preliminary, provide new insights toward building scalable and practical quantum communication infrastructures.

\begin{appendices}

\section{Proof of Lemma~\ref{lem:x01}}
\label{app:lem:x01}

We assume that $\ket{G_{n_1+1,n_2+1}}$ as equation~\eqref{eq:lem:x01} holds, and we must prove that $\ket{\bar{G}_{n_1,n_2}}$ can be obtained from $\ket{G_{n_1+1,n_2+1}}$. Additionally, we denote with $\overline{N}^i \eqdef \overline{N}(v_{1.i})$ the set of opposite remote nodes for node $v_{1.i}$ in the original graph $G_{n_1+1,n_2+1}$. As mentioned in Lem.~\ref{lem:x01}, there are two cases of $\ket{G_{n_1+1,n_2+1}}$. Let us prove Case I and then Case II. 

The proof constructively follows by performing Pauli-$X$ measurement on super-nodes $s_2, s_1$ in order. For more details about Pauli measurement please refer to \cite{MazCalCac-24-QCNC}.
\begin{itemize}
    \item[i)] Pauli-X measurement on the super-nodes $s_2$ with the arbitrary neighbor client-nodes $v_{1.i}$ as $k_0$\cite{CheIllCac-24-QCE}. It is equivalent to perform the sequence of graph operations $\tau_{v_{1.i}} \left( \tau_{s_2}\big(\tau_{v_{1.i}}(G_{n_1+1,n_2+1})\big)-s_2 \right)$.  This yields to $G'_{n_1+1,n_2}$ as:        
        \begin{align}
            \label{eq:app:lem:x01.3}
            & G'_{n_1+1,n_2}  = \tau_{v_{1.i}} \big( \tau_{s_2}(\tau_{v_{1.i}}(G_{n_1+1,n_2+1}) ) -s_2 \big)= \nonumber \\
            & =( V_1, V_2 \setminus \{s_2\},  
            (\{v_{1.i}\} \times V_1) 
            \cup  \nonumber \\
            & \cup \big( ( V_1 \setminus \{v_{1.i}\} ) \times \overline{N}(v_{1.i}) \big) 
            \cup  \big( \bigcup_{ \substack{ v_{1.j} \in V_1,\\ j \neq i} }  \{ v_{1.j} \} \times \overline{N}^j \big)  ) 
        \end{align}
    \item[ii)] Pauli-X measurement on the super-nodes $s_1$ by choosing again $v_{1.i}$ as $k_0$ (which belongs now to $N(s_1)$ as a consequence of the first Pauli-X measurement). It is equivalent to perform the sequence of graph operations $\tau_{v_{1.i}} \left( \tau_{s_1}\big(\tau_{v_{1.i}}(G'_{n_1+1,n_2})\big)-s_1 \right)$. This yields to $\bar{G}_{n_1,n_2}$:
        \begin{align}
            \label{eq:app:lem:x01.4}
            \tau_{v_{1.i}}(\tau_{s_1}(\tau_{v_{1.i}}(G'_{n_1+1,n_2}))- s_1) &= \big( V_1, V_2,  \big( \bigcup_{v_{1.j} \in V_1} \bar{E}(v_{1.j})\big) \big)  \nonumber \\
            &= \bar{G}_{n_1,n_2}
        \end{align}
\end{itemize}
Thus the proof follows.

The proof of Case II follows similarly to Case I, by setting $k_0 = v_{2.i}$, an arbitrary neighbor client-node in QLAN $V_2$.

\end{appendices}

\bibliographystyle{IEEEtran}
\bibliography{biblio.bib}

@article{Cal-17,
	title={Optimal routing for quantum networks},
	author={Caleffi, Marcello},
	journal={IEEE Access},
	volume={5},
	pages={22299--22312},
	year={2017},
	publisher={IEEE}
}

@article{PirDur-19,
	year = 2019,
	month = {mar},
	publisher = {{IOP} Publishing},
	volume = {21},
	number = {3},
	pages = {033003},
	author = {A Pirker and W Dür},
	title = {A quantum network stack and protocols for reliable entanglement-based networks},
	journal = {New Journal of Physics},
}

@inproceedings{ShoQia-20,
	author = {Shi, Shouqian and Qian, Chen}, 
	title = {Concurrent Entanglement Routing for Quantum Networks: Model and Designs},
	year = {2020}, 
	booktitle = {Proc. of ACM SIGCOMM '20},
	pages = {62–75},
	numpages = {14}
}

@article{IllCalMan-22,
    title={Quantum Internet Protocol Stack: a Comprehensive Survey},
    author={Illiano, Jessica and Caleffi, Marcello and Manzalini, Antonio and Cacciapuoti, Angela Sara},
    journal={Computer Networks},
    volume={213},
    year={2022}
}

@article{HeiDurEis-06,
  title={Entanglement in graph states and its applications},
  author={Hein, Marc and D{\"u}r, Wolfgang and Eisert, Jens and Raussendorf, Robert and Nest, M and Briegel, H-J},
  journal={arXiv preprint quant-ph/0602096},
  year={2006}
}

@article{CheIllCac-25,
      title={Entanglement-based artificial topology: Neighboring remote network nodes},
      author={Chen, Si-Yi and Illiano, Jessica and Cacciapuoti, Angela Sara and Caleffi, Marcello},
      journal={IEEE Open Journal of the Communications Society},
      year={2025},
      publisher={IEEE}
}

@article{CacIllCal-23,
  title={Quantum Internet Addressing},
  author={Cacciapuoti, Angela Sara and Illiano, Jessica and Caleffi, Marcello},
  journal={IEEE Network},
  year={2023},
  publisher={IEEE}
}

@article{MazCalCac-24-QCNC,
  title={{Quantum LAN: On-Demand Network Topology via Two-colorable Graph States}},
  author={Mazza, Francesco and Caleffi, Marcello and Cacciapuoti, Angela Sara},
  journal={2024 QCNC},
  year={2024},
  pages={127-134}
}

@INPROCEEDINGS{LiuLiWang-25,
  author={Liu, Maoli and Li, Zhuohua and Wang, Xuchuang and Lui, John C. S.},
  booktitle={IEEE INFOCOM 2024 - IEEE Conference on Computer Communications}, 
  title={LinkSelFiE: Link Selection and Fidelity Estimation in Quantum Networks}, 
  year={2024},
  volume={},
  number={},
  pages={1421-1430},
  doi={10.1109/INFOCOM52122.2024.10621263}}

@INPROCEEDINGS{LiuLiCai-24,
  author={Liu, Maoli and Li, Zhuohua and Cai, Kechao and Allcock, Jonathan and Zhang, Shengyu and Lui, John C.S.},
  booktitle={IEEE INFOCOM 2024 - IEEE Conference on Computer Communications}, 
  title={Quantum BGP with Online Path Selection via Network Benchmarking}, 
  year={2024},
  volume={},
  number={},
  pages={1401-1410},
  doi={10.1109/INFOCOM52122.2024.10621359}}

@article{MazCalCac-24,
  title={Intra-QLAN Connectivity via Graph States: Beyond the Physical Topology},
  author={Mazza, Francesco and Caleffi, Marcello and Cacciapuoti, Angela Sara},
  journal={IEEE Transactions on Network Science and Engineering},
  year={2025},
  publisher={IEEE}
}

@inproceedings{CheIllCac-24-QCE,
  title={Scaling Quantum Networks: Inter-QLANs Artificial Connectivity},
  author={Chen, Si-Yi and Illiano, Jessica and Cacciapuoti, Angela Sara and Caleffi, Marcello},
  booktitle={2024 IEEE International Conference on Quantum Computing and Engineering (QCE)},
  volume={1},
  pages={1980--1988},
  year={2024},
  organization={IEEE}
}

@article{CheCacCal-25,
      title={Entanglement-Enabled Connectivity Bounds for Quantum Network}, 
      author={SiYi Chen and Angela Sara Cacciapuoti and Marcello Caleffi},
      year={2025},
      number={submit/6359916},
      journal={arXiv},
      primaryClass={quant-ph} 
}

@INPROCEEDINGS{AliChe-22,
  author={Farahbakhsh, Ali and Feng, Chen},
  booktitle={IEEE INFOCOM 2022 - IEEE Conference on Computer Communications}, 
  title={Opportunistic Routing in Quantum Networks}, 
  year={2022},
  volume={},
  number={},
  pages={490-499},
  doi={10.1109/INFOCOM48880.2022.9796816}}

@article{CalAmoFer-24,
  title={Distributed quantum computing: a survey},
  author={Caleffi, Marcello and Amoretti, Michele and Ferrari, Davide and Illiano, Jessica and Manzalini, Antonio and Cacciapuoti, Angela Sara},
  journal={Computer Networks},
  volume={254},
  pages={110672},
  year={2024},
  publisher={Elsevier}
}

@article{WenJosSte-18,
  title={An entanglement-based wavelength-multiplexed quantum communication network},
  author={Wengerowsky, S{\"o}ren and Joshi, Siddarth Koduru and Steinlechner, Fabian and H{\"u}bel, Hannes and Ursin, Rupert},
  journal={Nature},
  volume={564},
  number={7735},
  pages={225--228},
  year={2018},
  publisher={Nature Publishing Group UK London}
}

@article{LiXueLi-23,
  title={Entanglement-assisted quantum networks: Mechanics, enabling technologies, challenges, and research directions},
  author={Li, Zhonghui and Xue, Kaiping and Li, Jian and Chen, Lutong and Li, Ruidong and Wang, Zhaoying and Yu, Nenghai and Wei, David SL and Sun, Qibin and Lu, Jun},
  journal={IEEE Communications Surveys \& Tutorials},
  volume={25},
  number={4},
  pages={2133--2189},
  year={2023},
  publisher={IEEE}
}

@article{MorDur-24,
  title={Influence of noise in entanglement-based quantum networks},
  author={Mor-Ruiz, Maria Flors and D{\"u}r, Wolfgang},
  journal={IEEE Journal on Selected Areas in Communications},
  year={2024},
  publisher={IEEE}
}

@misc{rfc9340,
    series =    {Request for Comments},
    number =    9340,
    howpublished =  {RFC 9340},
    publisher = {RFC Editor},
    doi =       {10.17487/RFC9340},
    author =    {Wojciech Kozlowski and Stephanie Wehner and Rodney Van Meter and Bruno Rijsman and Angela Sara Cacciapuoti and Marcello Caleffi and Shota Nagayama},
    title =     {{Architectural Principles for a Quantum Internet}},
    pagetotal = 37,
    year =      2023,
    month =     mar,
}

\end{document}